\documentclass[%
 aip,
 amsmath,amssymb,
 reprint,%
]{revtex4-1}

\usepackage{graphicx}% Include figure files
\usepackage{dcolumn}% Align table columns on decimal point
\usepackage{bm}% bold math
\usepackage[utf8]{inputenc}
\usepackage[T1]{fontenc}
\usepackage{mathptmx}
\usepackage{etoolbox}

\makeatletter
\def\@email#1#2{%
 \endgroup
 \patchcmd{\titleblock@produce}
  {\frontmatter@RRAPformat}
  {\frontmatter@RRAPformat{\produce@RRAP{*#1\href{mailto:#2}{#2}}}\frontmatter@RRAPformat}
  {}{}
}%
\makeatother
\begin{document}

\preprint{AIP/123-QED}

\title[Internal noise in convolutional neural networks]{Impact of internal noise on convolutional neural networks}
% Force line breaks with \\
\author{I.D. Kolesnikov}%
\author{N. Semenova}%
 \email{semenovani@sgu.ru}
 \affiliation{Saratov State University, Astrakhanskaya str. 83, Saratov 410012, Russia}%

\date{\today}% It is always \today, today,

\begin{abstract}
In this paper, we investigate the impact of noise on a simplified trained convolutional network. The types of noise studied originate from a real optical implementation of a neural network, but we generalize these types to enhance the applicability of our findings on a broader scale. The noise types considered include additive and multiplicative noise, which relate to how noise affects individual neurons, as well as correlated and uncorrelated noise, which pertains to the influence of noise across one layers. We demonstrate that the propagation of uncorrelated noise primarily depends on the statistical properties of the connection matrices. Specifically, the mean value of the connection matrix following the layer impacted by noise governs the propagation of correlated additive noise, while the mean of its square contributes to the accumulation of uncorrelated noise. Additionally, we propose an analytical assessment of the noise level in the network's output signal, which shows a strong correlation with the results of numerical simulations.
\end{abstract}

\maketitle

\begin{quotation}
Artificial neural networks (ANNs) have emerged as a powerful tool in recent years, addressing problems once deemed unsolvable without natural intelligence. Despite the existence of high-power computing clusters with the ability to parallelize computations, modeling a neural network on digital equipment is a bottleneck in network scaling, speed of receiving or processing information and energy efficiency. Recently, there has been a surge of interest among researchers in the development of hardware neural networks \cite{Karniadakis2021}, where neurons and their interconnections are realized as physical devices capable of learning and solving problems. These are often referred to in the literature as ``analog neural networks'' or ``hardware neural networks''. Unlike traditional simulations on computers, hardware ANNs represent tangible devices where the neurons and connections are implemented at the physical level, leading to significant improvements in processing speed and energy efficiency.  In hardware ANNs, multiple internal noise sources with varying properties can affect performance. Therefore, investigating the impact of different noise types on the operation of such networks, as well as exploring topologies that enable the network to mitigate internal noise, is a critical and relevant challenge. In this paper we study the impact of different noise types on convolutional neural networks.
\end{quotation}

\section{Introduction}\label{sec:intro}
Artificial neural networks (ANNs) have emerged as a powerful tool in recent years, addressing problems once deemed unsolvable without natural intelligence \cite{Lecun2015}. Today, neural networks are extensively utilized in various applications, including diagnostic systems, image recognition \cite{Krizhevsky2017,Maturana2015}, classification tasks, speech recognition \cite{Graves2013}, climate prediction \cite{Kar2009}, and much more. The foundational structure of ANNs was initially inspired by the human brain; however, modern ANNs exhibit neuron properties and connection features that are tailored to specific problems, diverging significantly from their biological counterparts.

%From a practical standpoint, modeling ANNs and executing computations on computers or computing clusters is a resource-intensive task. According to statistics from OpenAI \cite{openAI}, the performance growth of ANNs over the past two decades has accelerated dramatically. Prior to 2012, performance (measured in Petaflop/s-days) doubled approximately every two years. Since 2012, this doubling period has shortened to just 3-4 months. While advancements in computer hardware have progressed, they have not kept pace with this exponential growth in ANN performance. Consequently, simulating neural networks on digital platforms has become a bottleneck, affecting network scalability, information processing speed, and energy efficiency \cite{Hasler2013, Gupta2015}.

Recently, there has been a surge of interest among researchers in the development of hardware neural networks \cite{Karniadakis2021}, where neurons and their interconnections are realized as physical devices capable of learning and solving problems. These are often referred to in the literature as ``analog neural networks'' or ``hardware neural networks''. Unlike traditional simulations on computers, hardware ANNs represent tangible devices where the neurons and connections are implemented at the physical level, leading to significant improvements in processing speed and energy efficiency \cite{Aguirre2024, Chen2023}. This area has seen exponential growth in research focused on hardware ANNs, particularly those utilizing lasers \cite{Brunner2013a}, memristors \cite{Tuma2016}, and spin-transfer oscillators \cite{Torrejon2017}. Connection between neurons in optical implementations of ANNs leverages principles of holography \cite{Psaltis1990}, diffraction \cite{Bueno2018, Lin2018}, integrated Mach-Zehnder modulator networks \cite{Shen2017}, wavelength division multiplexing \cite{Tait2017}, and optical links produced via 3D printing \cite{Moughames2020, Dinc2020, Moughames2020a}.

%In hardware ANNs, the challenges associated with memory access and mathematical operations on large datasets are alleviated, as each neuron corresponds to a hardware nonlinear component, and each connection represents a physical communication channel. This architecture significantly enhances information processing speed and energy efficiency. 
At the same time, hardware ANNs are susceptible to internal noise generated by the components of these devices. Therefore, investigating the impact of different noise types on the operation of such networks, as well as exploring topologies that enable the network to mitigate internal noise, is a critical and relevant challenge. In previous studies, we examined the effects of internal noise on trained deep feedforward networks \cite{Semenova2022NN} and recurrent networks \cite{Semenova2024echo}, and proposed universal strategies for reducing internal network noise \cite{Semenova2022Chaos, Semenova2024Chaos}. This article focuses on another fundamentally important type of ANN —- convolutional neural networks. Following the approach of our earlier works, we will analyze additive and multiplicative noise based on their specific impact on neurons, as well as correlated and uncorrelated noise based on their effects on network layers composed of neurons.

Convolutional neural networks (CNNs) are a specialized subset of deep neural networks. The operation of a convolutional neural network is typically understood as a progression from specific image features to increasingly abstract representations, ultimately leading to the extraction of high-level concepts. During the training process, the network dynamically adjusts itself to create a necessary hierarchy of abstract features (sequences of feature maps), filtering out irrelevant details while emphasizing the essential ones. From a structural perspective, this results in alternating convolutional layers and pooling layers. These functional characteristics make convolutional networks highly effective for image recognition tasks, and they are widely applied in various fields, including computer vision and natural language processing \cite{Li2022}. The unique type of connections and the presence of layers with alternating topologies allow for a fundamentally new perspective on issues related to noise exposure and noise accumulation.

In this article, we examine the impact of internal noise at the convolutional layer stage of CNN and analyze how the variance of the network's output signal changes. We then introduce a pooling layer and investigate which method is more critical for noise accumulation: MaxPooling or MeanPooling. Additionally, the article presents an analytical assessment of noise influence, relying solely on the statistical properties of the connection matrices rather than on numerical modeling results.

\section{System under study}\label{sec:systemUnderStudy}
\subsection{Convolutional neural networks}\label{sec:systemUnderStudy:nets}

This paper studies the impact of noise on a convolutional neural network (CNN). In order to eliminate the complex impact of statistical characteristics of connection matrices, a simplified trained network consisting of the main components inherent in convolutional networks is considered. Previously, we have studied deep feedforward networks \cite{Semenova2022NN}, where conclusions were made about the propagation of noise based on the variance of the noise influence and the statistical properties of the connection matrices in the trained network. In the present article, we will provide the conclusions based on previous results, but adjusted according with the features of the convolutional and pooling layers inherent to CNNs.

The networks will be trained using a standard task of handwritten digit recognition from MNIST database \cite{LeCun1998}. This database contains 70,000 images of size 28$\times$28 pixels in a grayscale. Some of these images are used to train the network (60,000), and the rest are used for testing. When working with the MNIST database, certain conditions are imposed on the input and output layers of the network. The input layer must be designed in such a way that each input neuron receives the value of the corresponding pixel in the image as input. Since the images are 28$\times$28 pixels in size, the input layer must consist of 784 neurons. For ease of use, the values obtained from the input image will be normalized by 255 so that the network's input values belong to the range $[0,1]$.

The neural network must be trained to solve the classification task, so each input image must be assigned to one of 10 possible classes (digits 0--9). Then the output layer must consist of 10 neurons, with each output neuron responsible for its own digit. The output signal of the network is not the output signal of the output neurons themselves, but which output neuron has a maximum value. For example, if an image with the digit 0 is transmitted to the network input, then the 0th output neuron must have the maximum output value. This operation is called softmax.

\begin{figure}[t]
\includegraphics[width=\linewidth]{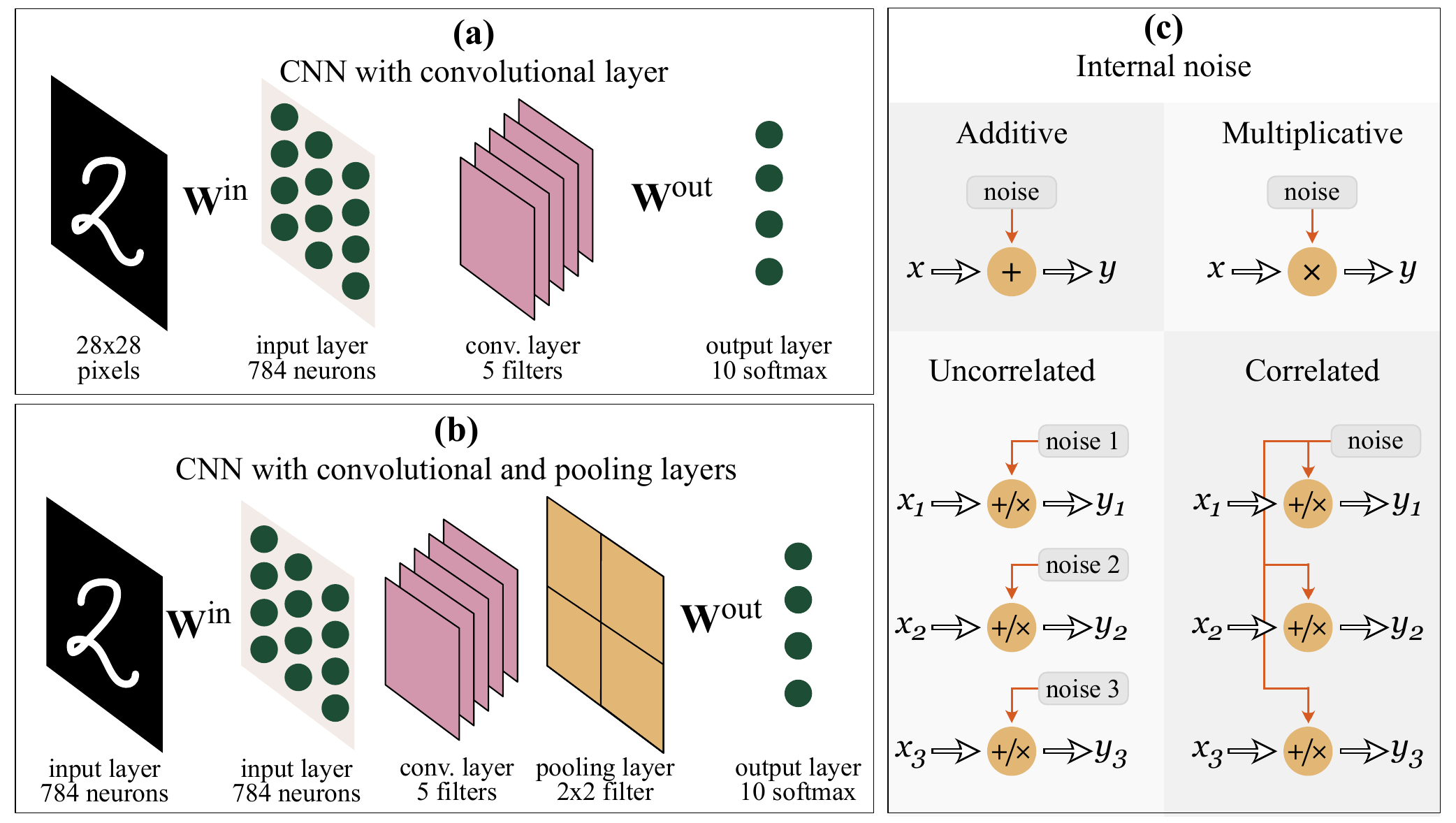}
\caption{\label{fig:scheme}Schematic representation of considered neural networks (a),(b) and methods of introducing the noise exposure (c).}
\end{figure}

\textbf{Convolutional layer.} In the following sections, the noise will be added to several CNNs, gradually increasing its complexity and introducing new network components. Fig.~\ref{fig:scheme} schematically shows all the gradual complications of the network. First, we consider the network with one convolutional layer (Fig.~\ref{fig:scheme}(a)). The purpose of the convolution layer is to filter the image. This layer includes one or several filters with a kernel of a certain size. From mathematical point of view the convolution layer is a matrix or several matrices (if there are several filters) of a certain size, the values of which are selected during the training process. The application of a filter of size 3$\times$3 at the convolution stage is as follows. For example, from the input layer we received a matrix of values consisting of 784 values. For ease of interpretation, we arrange them so that they look like a matrix of size 28$\times$28. The filter is applied to the upper left corner of the matrix, capturing 9 elements, then these 9 matrix elements and the filter elements are element-by-element multiplied, and their total value is written to a new matrix. Then the filter is shifted one column to the right, and the procedure is repeated until the filter reaches the end of the matrix row; next, the filter is shifted one row down and the whole procedure is repeated again. At the end, after applying the filter, we get a new matrix, but its size is already 26$\times$26 (i.e. 676 neurons will be needed to process these values).

There may be several such filters at the convolution layer. If there is only one 3$\times$3 filter, then there are 676 neurons left after the convolution stage, and the output matrix $\mathbf{W}^\mathrm{out}$ connecting the convolutional layer with the output layer has a size of 676$\times$10. If two filters are used, then there are 1352 neurons, and the size of $\mathbf{W}^\mathrm{out}$ is 1352$\times$10. Using 5 filters already leads to $676*5=3380$ neurons, and so on. For the network topology shown in Fig.~\ref{fig:scheme},~\textit{a}, noise will be introduced into the neurons obtained as a result of applying the convolution layer. As an example, in this paper we show the results for 5 filters of size 3$\times$3 in convolutional layer. We have considered the other combinations, but the overall qualitative results in terms of noise accumulation were the same.  
 
\textbf{Pooling layer.} The pooling layer in CNNs is usually used to reduce the dimensionality that is created after convolution stage (Fig.~\ref{fig:scheme}(b)). The role of pooling is to select one output value of one neuron from a certain group of neurons. The transformation has the form of non-intersecting rectangles or squares of the same size, each of which captures a certain group of neurons for the subsequent transformation of their values into one. The most commonly used is the selection of the maximum value (MaxPooling) or the average value (MeanPooling).
Pooling is interpreted as follows: if some features have already been identified in the previous convolution operation, then such a detailed image is no longer needed for further processing, and it is compacted to a less detailed one. For example, after applying convolution with one 3$\times$3 filter, 676 neurons are formed, which can be arranged in a 26$\times$26 square. Applying downsampling with a 2$\times$2 filter results in the 26$\times$26 matrix of values being split into 2$\times$2 cells. Only one value (maximum or average) is selected from each cell, and thus the matrix dimension is reduced to 13$\times$13, which corresponds to 169 neurons, then the size of the output matrix $\mathbf{W}^\mathrm{out}$ already becomes 169$\times$10. 

In this paper, we consider CNNs with five 3$\times$3 filters in convolutional layer, resulting in $676\cdot 5=3380$ neurons after convolutional stage. Pooling with filter 2$\times$2 leads to $169\cdot 5=845$ neurons before the output layer.
 
\subsection{Noise types}
The properties and features of introducing the internal noise are similar to our previous works \cite{Semenova2022NN, Semenova2022Chaos}. This allows to compare and underline the common features and difference between noise propagation in feedforward and convolutional neural networks. The original types of internal noise, their intensities and introduction methods were obtained from the hardware implementation of the ANN in the optical experiment proposed in Ref.~\cite{Bueno2018}. Here we will consider different noise intensities to make the results more general and applicable to other hardware networks.

All types of noise under consideration are schematically shown in Fig.~\ref{fig:scheme}(c). Depending on how noise affects the signal of one individual neuron, additive and multiplicative noise will be considered. We will assume that each neuron has its own noise-free output signal $x_i$ before noise influence including the impact of the connection matrices, previous layers and all additional operations such as convolution or pooling. Then the noise influences are introduced into this signal:

\begin{equation} \label{eq:noiseTypes}
y_i(t) = x_i(t)\cdot \Big(1+\sqrt{2D_M}\xi_M(t,i))\Big) + \sqrt{2D_A}\xi_A(t,i).
\end{equation}
Additive noise $\xi_A$ is added to the noise-free output signal, and multiplicative noise (with indices ``M'') is multiplied by it. The notation $\xi$ corresponds to white Gaussian noise with zero mean and unity variance. The multiplier $\sqrt{2D}$ is often called as the intensity of the noise influence, and it determines the total variance (dispersion) of the noise exposure. In the equation (\ref{eq:noiseTypes}), the index $i$ corresponds to the neuron number within one layer, and $t$ to the input image number.

The noise influences that are the same for groups of neurons can also be observed in hardware neural networks. Therefore, along with the classification of noise depending on the effect on a single neuron, it is also necessary to introduce a classification of noise effects depending on the effect on a group of neurons (in this article, this is one layer). The noise effect, the values of which are different for each new input image, but they are the same for all neurons within one layer, we will call correlated noise ($\sqrt{2D^C_A}\xi^C_A(t)$, $\sqrt{2D^C_M}\xi^C_M(t)$), while the noise exposure being different for these neurons, we will call uncorrelated noise ($\sqrt{2D^U_A}\xi^U_A(t,i)$, $\sqrt{2D^U_M}\xi^U_M(t,i)$).

Thus, in total, four types of noise exposure are considered in the article:
\begin{itemize}
\item additive uncorrelated noise: \\$y_i(t) = x_i(t) + \sqrt{2D^U_A}\xi^U_A(t,i)$;
\item additive correlated noise: \\$y_i(t) = x_i(t) + \sqrt{2D^C_A}\xi^C_A(t)$;
\item multiplicative uncorrelated noise: \\$y_i(t) = x_i(t)\cdot \big(1+\sqrt{2D^U_M}\xi^U_M(t,i)\big)$;
\item multiplicative correlated noise: \\$y_i(t) = x_i(t)\cdot \big(1+\sqrt{2D^C_M}\xi^C_M(t)\big)$;
\end{itemize}

\subsection{Noise estimation in numerical simulation}\label{sec:systemUnderStudy:calc}

Further, we will show how different noise influences change the variance of the output signal depending on its mean. For example, see Fig.~\ref{fig:conv_uncorr}. These dependencies are prepared in the next way. Each input image is repeated $K=100$ times, then the mean value and variance are averaged over $K$ repetitions for each input image and each output neuron. Therefore, each panel in Fig.~\ref{fig:conv_uncorr}(a) contains 100,000 points according to 10,000 testing images and 10 output neurons. Comparing the panels (a) and (c) one can see, that in some cases we used several colors, and sometimes only one. This is because sometimes the variance strongly depend on which of the 10 output neurons the output signal is read from. The color corresponds to the ordinal number of the output neuron. In cases where data from different output neurons overlapped, only one color was used.

\section{Results}\label{sec:results}
\subsection{Noise in CNN with convolutional layer}\label{sec:results:conv}
\begin{figure}[t]
\includegraphics[width=\linewidth]{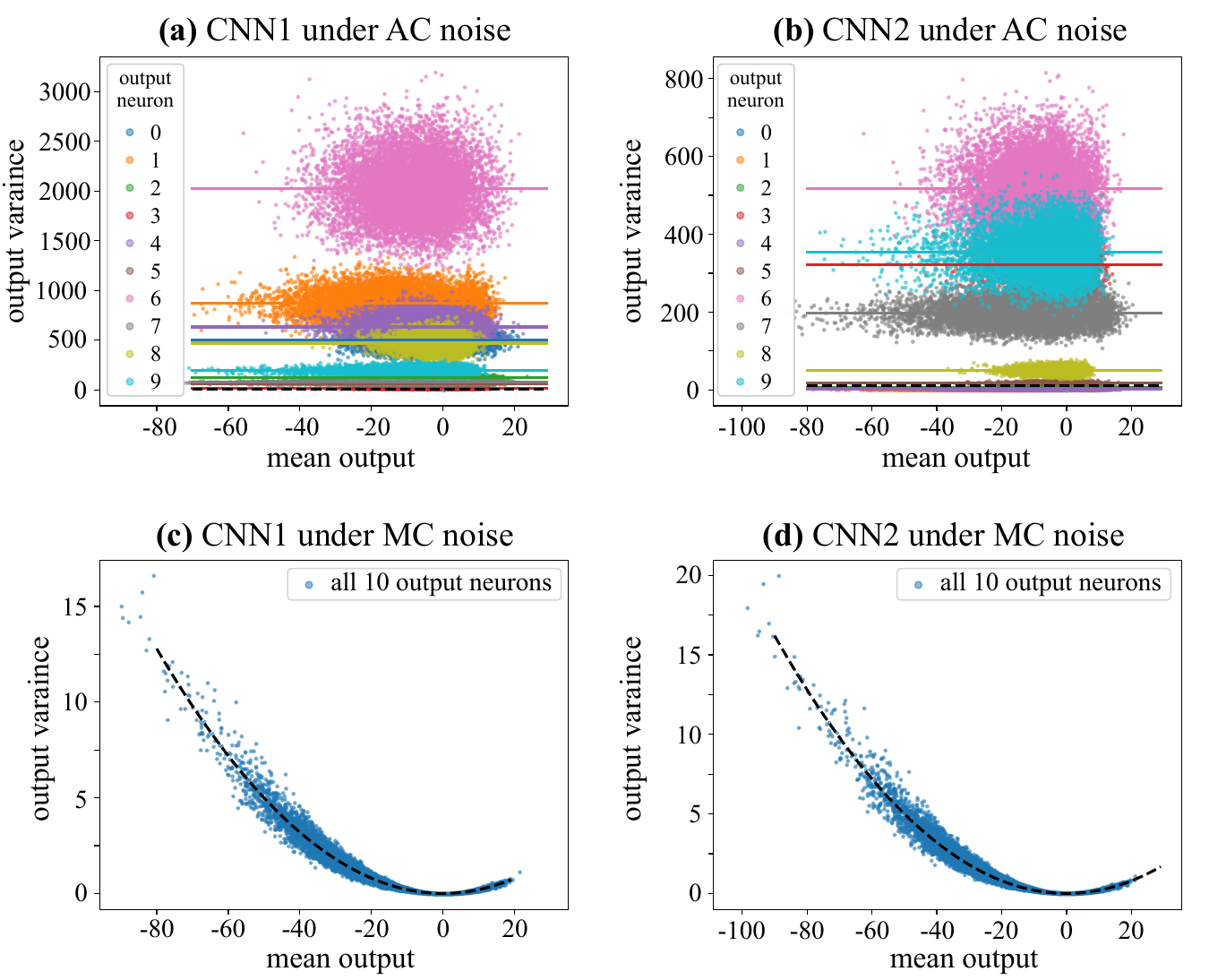}
\caption{\label{fig:conv_corr} The impact of correlated noise on two trained CNNs. All panels contain the dependencies of variance (dispersion) on mean of the output signal from each of 10 output neurons. Considered noise types: (a,b) -- additive correlated noise with $D^C_A=10^{-3}$, (c,d) -- multiplicative correlated noise with $D^C_M=10^{-3}$. The results of numerical simulation are shown by points, while the lines of different colors correspond to analytical estimation of the noise level based on (\ref{eq:varXout_final2}) with averaging over one of 10 rows of $\mathbf{W}^\mathrm{out}$, black dashed lines were prepared using averaging over the entire matrix.}
\end{figure}

In this section, we consider how the internal noise in the convolutional layer affects the accuracy of trained network. Considered network schematically shown in Fig.~\ref{fig:scheme}(a) has a convolutional layer with five filters of size 3$\times$3 and without pooling. We have considered also other combinations in convolutional layer: 1 filter of size 3$\times$3, 1 filter 5$\times$5, 2 filters 3$\times$3, but qualitatively it led to the same results. Noise was introduced into the already trained network into neurons after the convolution stage (3380 neurons). Figure~\ref{fig:conv_corr} shows the variance of CNNs' output signal in the case of additive (a,b) and multiplicative (c,d) correlated noise for two trained CNNs. CNN1 has a final training accuracy 94.13\% and testing accuracy 92.32\% while CNN2 has training accuracy 93.88\% and testing accuracy 92.11\% Their accuracies are almost identical, but comparing the range of the obtained output variance values, one can see that they differ by orders of magnitude. Although all the graphs were prepared for the same noise intensity of $0.001$. For this reason, we have chosen these two networks for demonstration how different types of noise can be accumulated depending on connection matrices of trained network.

The dependences of variance on mean output signal obtained for additive correlated noise (Fig.~\ref{fig:conv_corr}(a,b)) are shown in different colors according with the output neuron from which the output signal was taken. The procedure for these calculations was described in more detail in Sect.~\ref{sec:systemUnderStudy:calc}. In the case of multiplicative correlated noise, there is almost no difference between output neurons. The points obtained for different output neurons intersect strongly and overlap each other, therefore, Fig.~\ref{fig:conv_corr}(c,d) are built in the same color.

\begin{figure}[t]
\includegraphics[width=\linewidth]{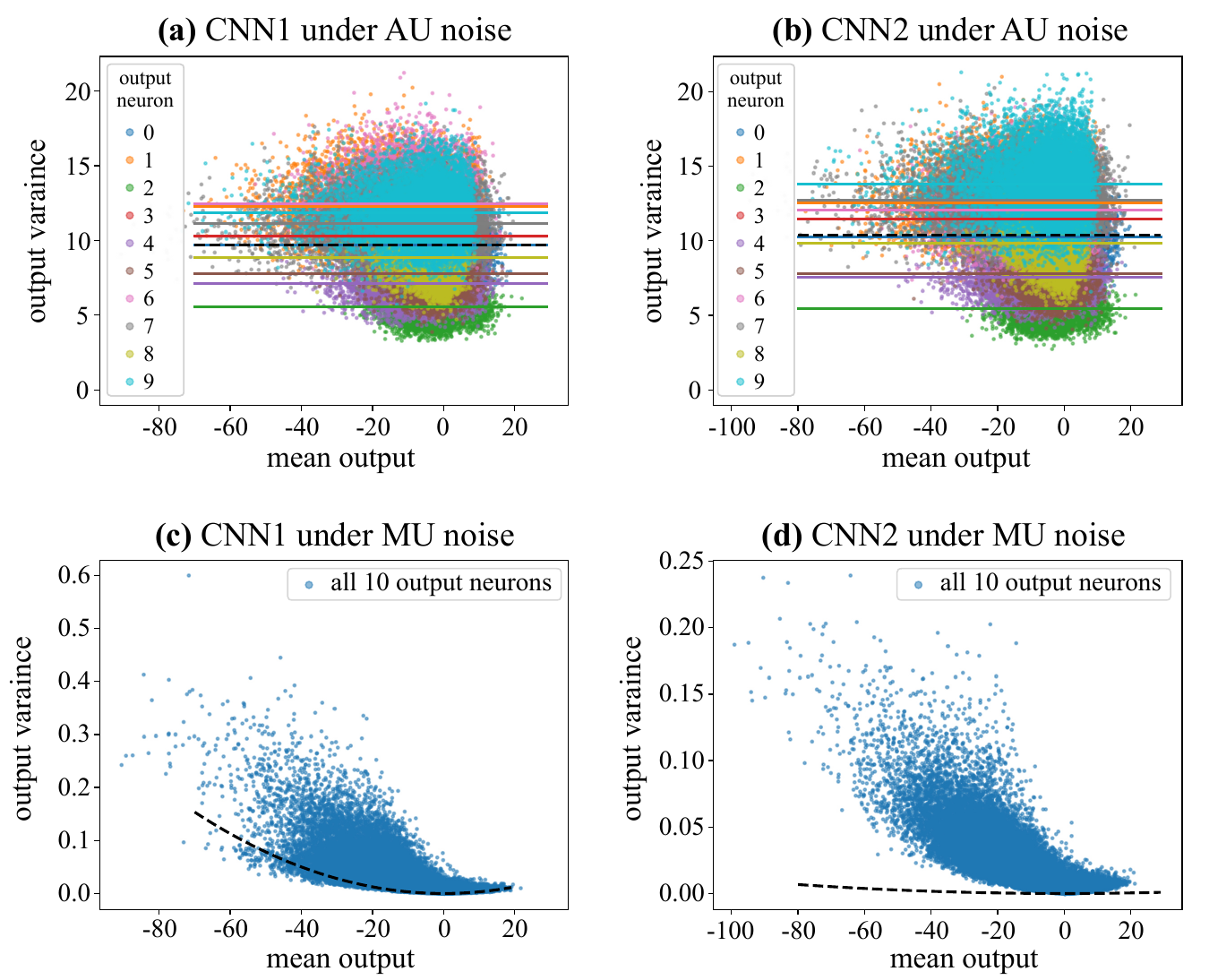}
\caption{\label{fig:conv_uncorr} The impact of uncorrelated noise on two trained CNNs. All panels contain the dependencies of variance (dispersion) on mean of the output signal from each of 10 output neurons. Considered noise types: (a,b) -- additive uncorrelated noise with $D^U_A=10^{-3}$, (c,d) -- multiplicative uncorrelated noise with $D^U_M=10^{-3}$. The results of numerical simulation are shown by points, while the lines of different colors correspond to analytical estimation of the noise level based on (\ref{eq:varXout_final2}) with averaging over one of 10 rows of $\mathbf{W}^\mathrm{out}$, black dashed lines were prepared using averaging over the entire matrix.}
\end{figure}

Similar calculations were prepared also for uncorrelated noise (Fig.~\ref{fig:conv_uncorr}). As can be seen from Fig.~\ref{fig:conv_uncorr}(a,b), the difference between the variance obtained from different output neurons became not so noticeable than for correlated additive noise.

It is important to note that Figures \ref{fig:conv_uncorr} and \ref{fig:conv_corr} contain not only points but also lines of corresponding colors. These lines show our theoretical predictions of variance level of the network output signal based on the properties of the network and the output connection matrix. More details about this will be discussed below.

\subsection{Analytical estimation of the output noise level}\label{sec:results:analytics}

Based on the convolution procedure described in Sect.~\ref{sec:systemUnderStudy:nets}, and the fact that there are 5 filters of size 3$\times$3 in the convolutional layer, we obtain that after convolution $N=3380$ neurons are formed. Therefore, the output connection matrix $\mathbf{W}^\mathrm{out}$ has the size $3380\times 10$, and the length of the vector $\mathbf{y}^\mathrm{conv}$ describing the output signal of neurons in convolutional layer is $N=3380$.

The output singal of output neurons of CNN in Fig.~\ref{fig:scheme} consists of output connection matrix $\mathbf{W}^\mathrm{out}$ and output signals of neurons from the convolutional layer after noise influence $\mathbf{y}^\mathrm{conv}$:
\begin{equation}\label{eq:xOut}
\begin{array}{c}
y^\mathrm{out}_i = \sum\limits^{N}_{j=0} W^\mathrm{out}_{ij} y^\mathrm{conv}_j, \ \ \text{where} \\
y^\mathrm{conv}_j = x^\mathrm{conv}_j \cdot\Big( 1+\sqrt{2D^C_M}\xi^C_M \Big)\Big( 1+\sqrt{2D^U_M}\xi^U_M(j) \Big) + \\
\sqrt{2D^C_A}\xi^C_A + \sqrt{2D^U_A}\xi^U_A(j),
\end{array}
\end{equation} 
where $x^\mathrm{conv}_j$ is the noise-free output of $j$th neuron from convolutional layer. In this equation, $y^\mathrm{conv}_j$ is its output after the noise impact including all four possible noise sources. All variables $y^\mathrm{out}_i$, $y^\mathrm{conv}_j$, $x^\mathrm{conv}_j$ and all $\xi$-variables also have the dependency on the input image number $t\in[0,10000)$ and the repetition number $k\in[0,K)$, but these indices have been omitted to make the equations easier to understand.

In the following, we will operate with the mathematical terms of variance (dispersion) $\mathrm{Var}[X]$ and expected value (mean value) $\mathrm{E}[X]$ of a random variable X.

According to general rules of mathematical operations for uncorrelated random variables $X$ and $Y$, one obtains $\text{Var}[X+Y]=\text{Var}[X]+\text{Var}[Y]$ and $\text{Var}[X\cdot Y]=\text{Var}[X]\cdot\big(\text{E}[Y]\big)^2+\Big( \big(\text{E}[X]\big)^2+\text{Var}[X] \Big)\cdot\text{Var}[Y]$. 
When $X$ is multiplied on a constant $C$, the variance becomes $\text{Var}[C\cdot X]=C^2\text{Var}[X]$. 
All considered noise sources $\xi$ have zero expected value and a variance equal to $1$. 
The final variance of the noise source is controlled by the corresponding noise intensity, and $\text{Var}(\sqrt{2D^U_A}\xi^{U,A}_{n,i})=2D^U_A$. We did a similar evaluation for the deep feedforward network in the paper \cite{Semenova2022NN}.

Taking the above into account, the variance of CNN's output $y^\mathrm{out}_i$ can be calculated as:
\begin{equation}\label{eq:varXout}
\begin{array}{c}
\mathrm{Var}[y^\mathrm{out}_i] = \mathrm{Var}\Big[\sum\limits^{j<N}_{j=0} W^\mathrm{out}_{ij} y^\mathrm{conv}_j\Big] = \\
\mathrm{Var}\bigg[ \sqrt{2D^C_A}\xi^C_A\cdot\sum\limits^{j<N}_{j=0} W^\mathrm{out}_{ij} + \sqrt{2D^U_A}\cdot\sum\limits^{j<N}_{j=0} W^\mathrm{out}_{ij}\xi^U_A(j) + \\
\Big(1+\sqrt{2D^C_M}\xi^C_M\Big)\cdot \sum\limits^{j<N}_{j=0} W^\mathrm{out}_{ij} x^\mathrm{out}_j \Big(1+\sqrt{2D^U_M}\xi^U_M(j)\Big) \bigg].
\end{array}
\end{equation}
As stated above, we assume that all noise sources $\xi$ have zero mean and unity variances, and that the variance of noise exposure is controlled by the noise intensity. Thus, the variance of $\sqrt{2D}\xi$ is $2D$. Therefore, the expression above can be rewritten as follows
\begin{equation}\label{eq:varXout_final1}
\begin{array}{c}
\mathrm{Var}[y^\mathrm{out}_i] = 2D^C_A\cdot\Big( \sum\limits^{j<N}_{j=0} W^\mathrm{out}_{ij} \Big)^2 + 2D^U_A\cdot \sum\limits^{j<N}_{i=0} \big( W^\mathrm{out}_{ij} \big)^2 + \\
2D^C_M\cdot\Big(E[x^\mathrm{out}_i]\Big)^2 + 2D^U_M(1+2D^C_M)\cdot\sum\limits^{j<N}_{j=0}\Big( W^\mathrm{out}_{ij} x^\mathrm{conv}_j  \Big)^2 .
\end{array}
\end{equation}
In Eq.~(\ref{eq:varXout_final1}), the sums $\sum\limits^{j<N}_{j=0} W^\mathrm{out}_{ij}$ and $\sum\limits^{j<N}_{i=0} \big( W^\mathrm{out}_{ij} \big)^2$ can be rewritten using mean and mean square values as:
\begin{equation}\label{eq:sums}
\begin{array}{c}
\sum\limits^{j<N}_{j=0} W^\mathrm{out}_{ij} = N\mu(\mathbf{W}^\mathrm{out}_i), \\
\sum\limits^{j<N}_{j=0} \big(W^\mathrm{out}_{ij}\big)^2 = N\eta(\mathbf{W}^\mathrm{out}_i),
\end{array}
\end{equation}
where $\eta(\mathbf{W}^\mathrm{out}_i)$ is the mean square, and $\mu(\mathbf{W}^\mathrm{out}_i)$ is the mean value of the $i$th raw of matrix $\mathbf{W}^\mathrm{out}$. We will also use the square of this value, denoted as $\mu^2(\mathbf{W}^\mathrm{out}_i)$. Therefore, Eq.~\ref{eq:varXout_final1} can be rewritten as
\begin{equation}\label{eq:varXout_final2}
\begin{array}{c}
\mathrm{Var}[y^\mathrm{out}_i] = 2D^C_A\cdot\Big(N\mu(\mathbf{W}^\mathrm{out}_i)\Big)^2 + 2D^U_A\cdot N\eta(\mathbf{W}^\mathrm{out}_i) + \\
2D^C_M\cdot\Big(E[y^\mathrm{out}_i]\Big)^2 + 2D^U_M(1+2D^C_M)\cdot\sum\limits^{j<N}_{j=0}\Big( W^\mathrm{out}_{ij} x^\mathrm{conv}_j  \Big)^2 .
\end{array}
\end{equation}
This equation is given in general form when all types of noise exists in the network. Variances in the output signal for individual types of noise can be obtained as follows.
\begin{itemize}
\item additive correlated noise: \\$\mathrm{Var}[y^\mathrm{out}_i] = 2D^C_A\cdot\Big(N\mu(\mathbf{W}^\mathrm{out}_i)\Big)^2$;
\item multiplicative correlated noise: \\$\mathrm{Var}[y^\mathrm{out}_i] = 2D^C_M\cdot\Big(E[y^\mathrm{out}_i]\Big)^2$;
\item additive uncorrelated noise: \\$\mathrm{Var}[y^\mathrm{out}_i] = 2D^U_A\cdot N\eta(\mathbf{W}^\mathrm{out}_i) $;
\item multiplicative uncorrelated noise: \\$\mathrm{Var}[y^\mathrm{out}_i] = 2D^U_M\cdot\sum\limits^{j<N}_{j=0}\Big( W^\mathrm{out}_{ij} x^\mathrm{conv}_j  \Big)^2$.
\end{itemize}
The lines depicted in Figs.~\ref{fig:conv_uncorr}, \ref{fig:conv_corr} were obtained using Eq.~\ref{eq:varXout_final2}. Both correlated noise influences and additive uncorrelated noise are in a good agreement with (\ref{eq:varXout_final2}). The impact of multiplicative uncorrelated noise is hard to predict (see Fig.~\ref{fig:conv_uncorr}(c,d)) due to the multiplier $\Big( W^\mathrm{out}_{ij} x^\mathrm{conv}_j  \Big)^2$ which is hard to approximate without modeling. In Fig.~\ref{fig:conv_uncorr}(c,d), we used the function $g(x)=2D^U_M \mu^2(\mathbf{W}^\mathrm{out}) \cdot x^2 $ to evaluate the minimal variance level.

The dependency of variance on the mean of the output signal of CNN with correlated multiplicative noise (Fig.~\ref{fig:conv_corr}(c,d)) can be easily approximated using Eq.~\ref{eq:varXout_final2} and function $g(x)=2D^C_M\cdot x^2$.

According to (\ref{eq:varXout_final2}), additive correlated and uncorrelated noise lead to variance which does not depend on the output signal, and its level is determined by statistical characteristics of output connection matrix $\mathbf{W}^\mathrm{out}$. 

As can be seen from the graphs, the proposed variants of theoretical noise level estimation correlate well with the results of numerical modeling. In order to understand the reason for such a large difference between the variatnces in CNN1 and CNN2, let us consider in more detail the statistical characteristics of the connection matrices in these networks. The mean, mean square and multipliers for variances in case of additive correlated $\Big(N\mu(\mathbf{W}^\mathrm{out}_i)\Big)^2$ and additive uncorrelated noise $N\eta(\mathbf{W}^\mathrm{out}_i)$ are given in Table~\ref{tab:CNN1} for CNN1 and in Table~\ref{tab:CNN2} for CNN2. 

\begin{table}
\caption{\label{tab:CNN1}Statistical characteristics of connection matrix $\mathbf{W}^\mathrm{out}$ in trained CNN1.}
\begin{ruledtabular}
\begin{tabular}{ccccc}
$i$ & $\mu(\mathbf{W}^\mathrm{out}_i)$ & $\eta(\mathbf{W}^\mathrm{out}_i)$ & $\Big(N\mu(\mathbf{W}^\mathrm{out}_i)\Big)^2$ & $N\eta(\mathbf{W}^\mathrm{out}_i)$ \\
\hline
0	&	0.149	&	1.438	&	252453.428	&	4860.723 \\
1	&	0.195	&	1.813	&	434192.320	&	6126.211 \\
2	&	0.073	&	0.822	&	60771.781	&	2777.127 \\
3	&	0.027	&	1.520	&	8309.023	&	5136.942 \\
4	&	0.166	&	1.051	&	316110.204	&	3551.828 \\
5	&	0.050	&	1.154	&	28736.982	&	3899.119 \\
6	&	0.298	&	1.844	&	1010896.175	&	6231.526 \\
7	&	0.059	&	1.645	&	39026.439	&	5558.328 \\
8	&	0.144	&	1.314	&	235407.817	&	4440.406 \\
9	&	0.092	&	1.753	&	97085.073	&	5924.392 \\
\hline
averaged	&	0.125	&	1.435	&	248298.924	&	4850.660
\\
\end{tabular}
\end{ruledtabular}
\end{table}

\begin{table}
\caption{\label{tab:CNN2} Statistical characteristics of connection matrix $\mathbf{W}^\mathrm{out}$ in trained CNN2.}
\begin{ruledtabular}
\begin{tabular}{ccccc}
$i$ & $\mu(\mathbf{W}^\mathrm{out}_i)$ & $\eta(\mathbf{W}^\mathrm{out}_i)$ & $\Big(N\mu(\mathbf{W}^\mathrm{out}_i)\Big)^2$ & $N\eta(\mathbf{W}^\mathrm{out}_i)$ \\
\hline
0	&	0.015	&	1.518	&	2502.034	&	5129.647 \\
1	&	0.008	&	1.859	&	791.007	&	6283.185 \\
2	&	0.018	&	0.809	&	3603.328	&	2733.674 \\
3	&	-0.119	&	1.693	&	161337.140	&	5721.468 \\
4	&	-0.009	&	1.115	&	950.052	&	3767.272 \\
5	&	-0.028	&	1.154	&	8846.566	&	3899.028 \\
6	&	0.150	&	1.786	&	258405.756	&	6035.872 \\
7	&	-0.093	&	1.879	&	98896.633	&	6349.850 \\
8	&	-0.047	&	1.455	&	25394.271	&	4918.638 \\
9	&	-0.124	&	2.040	&	176742.159	&	6896.120 \\
\hline
averaged	&	-0.023	&	1.531	&	73746.895	&	5173.476
\\
\end{tabular}
\end{ruledtabular}
\end{table}

The largest spread in values is seen for $\Big(N\mu(\mathbf{W}^\mathrm{out}_i)\Big)^2$ depending on the averaging over the $i$th row of the matrix $\mathbf{W}^\mathrm{out}$. This multiplier refers to the variance of the output signal in the case of additive correlated noise. This explains why in Fig.~\ref{fig:conv_corr}(a,b) we see such a difference between the variances obtained from different output neurons.

\subsection{Noise in CNN with convolutional and pooling layers}\label{sec:results:pool}
In this section, we consider a convolutional network consisting of a convolution layer and a pooling layer, schematically shown in Fig.~\ref{fig:scheme},~\textit{b}.

Here we show the results for two trained networks with MaxPooling and MeanPooling after convolutional layer. We have trained several networks of each type, but the results were qualitatively the same. The considered CNN with MaxPooling has the training accuracy 99.43\% and testing accuracy 96.65\%, while the CNN with MeanPooling has training accuracy 93.54\% and testing accuracy 92.66\%.

\begin{figure}[t]
\includegraphics[width=\linewidth]{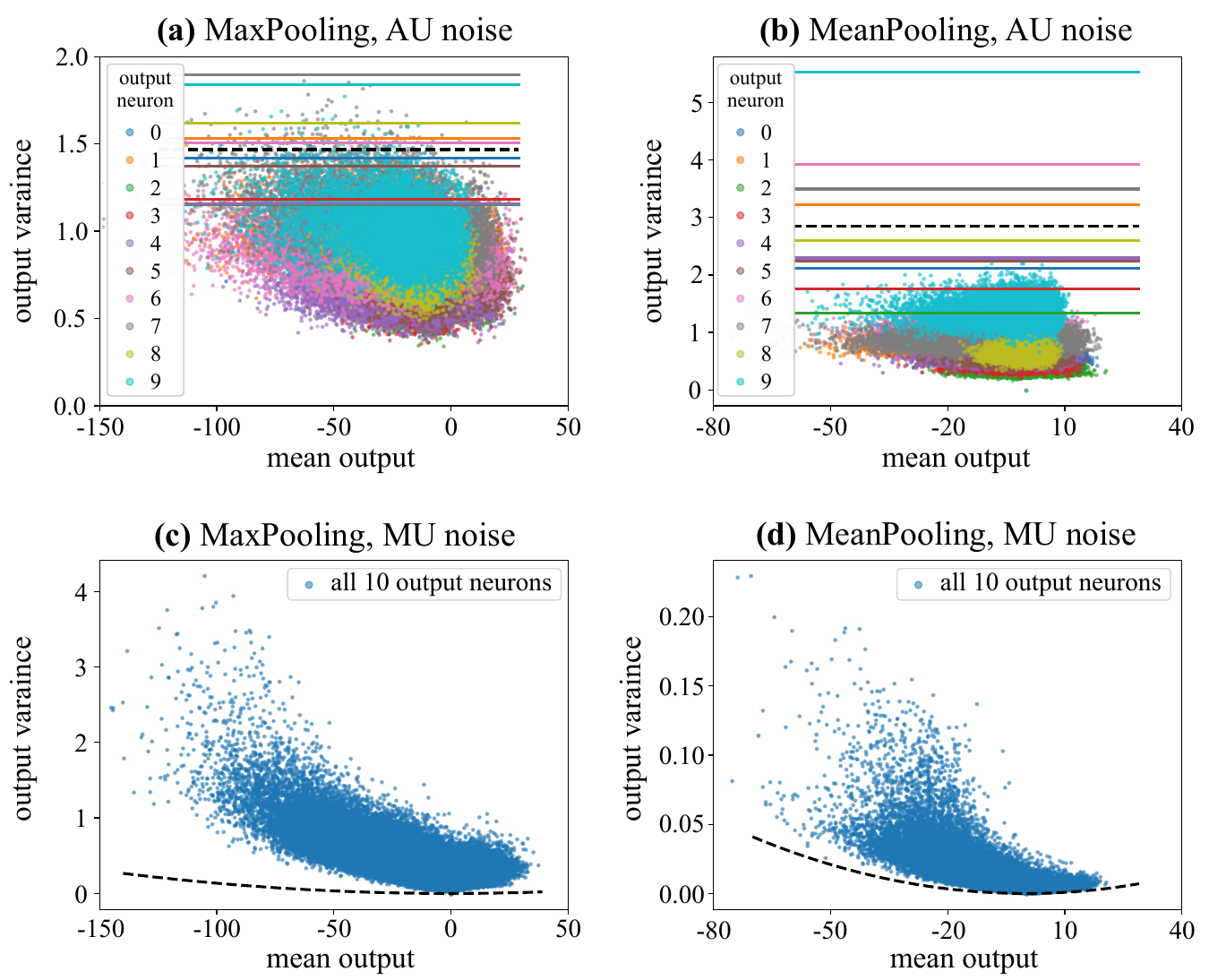}
\caption{\label{fig:pool_uncorr} The impact of uncorrelated noise on the output of trained CNNs with MaxPooling (a,c) and MeanPooling (b,d). All panels contain the dependencies of variance (dispersion) on mean of the output signal from each of 10 output neurons. Considered noise types: (a,b) -- additive uncorrelated noise with $D^U_A=10^{-3}$, (c,d) -- multiplicative uncorrelated noise with $D^U_M=10^{-3}$. The results of numerical simulation are shown by points, while the lines of different colors correspond to analytical estimation of the noise level based on (\ref{eq:varXout_final2}) with averaging over one of 10 rows of $\mathbf{W}^\mathrm{out}$, black dashed lines were prepared using averaging over the entire matrix.}
\end{figure}

Figure~\ref{fig:pool_uncorr} shows the impact of uncorrelated noise on CNNs with MaxPooling layer (a,c) and MeanPooling layer (b,d) after convolutional layer of the same configuration as in previous section. The lines correspond to theoretical prediciton of the noise level in absence of pooling layer. This allows to compare what we would get if there were no pooling layers.

Figure~\ref{fig:pool_uncorr}(b,d) shows that both uncorrelated noise types can be significantly reduced by using MeanPooling in the pooling layer. This is a fairly logical result, since in the article \cite{Semenova2024Chaos} we proposed a pooling technique to reduce the influence of uncorrelated noise. The essence of this method was to create duplicate neurons and then average them. In fact, this is very similar to the procedure that occurs when using a pooling layer with averaging (MeanPooling). 

In the case of MaxPooling (Fig.~\ref{fig:pool_uncorr}(a)), the situation is not so clear. Both MeanPooling and MaxPooling reduces the impact of additive uncorrelated noise. New levels of variance are much lower than variance without pooling (Fig.~\ref{fig:conv_uncorr}(a,b)). Comparing the range of variance for multiplicative uncorrelated noise in Fig.~\ref{fig:pool_uncorr}(c,d), one can see that the variance for MaxPooling is much higher. This is caused by the increasement of output values of CNN due to MaxPooling. According with our analytical predictions the output variance in this case is directly proportional to the output of CNN. MaxPooling increases of output values and therefore increases the variance in the case of multiplicative uncorrelated noise. The lines in Fig.~\ref{fig:pool_uncorr} correspond to analytical prediction of the noise level in the same trained CNNs (with the same statistical characteristics of $\mathbf{W}^\mathrm{out}$) but without pooling layer.

\begin{figure}[t]
\includegraphics[width=\linewidth]{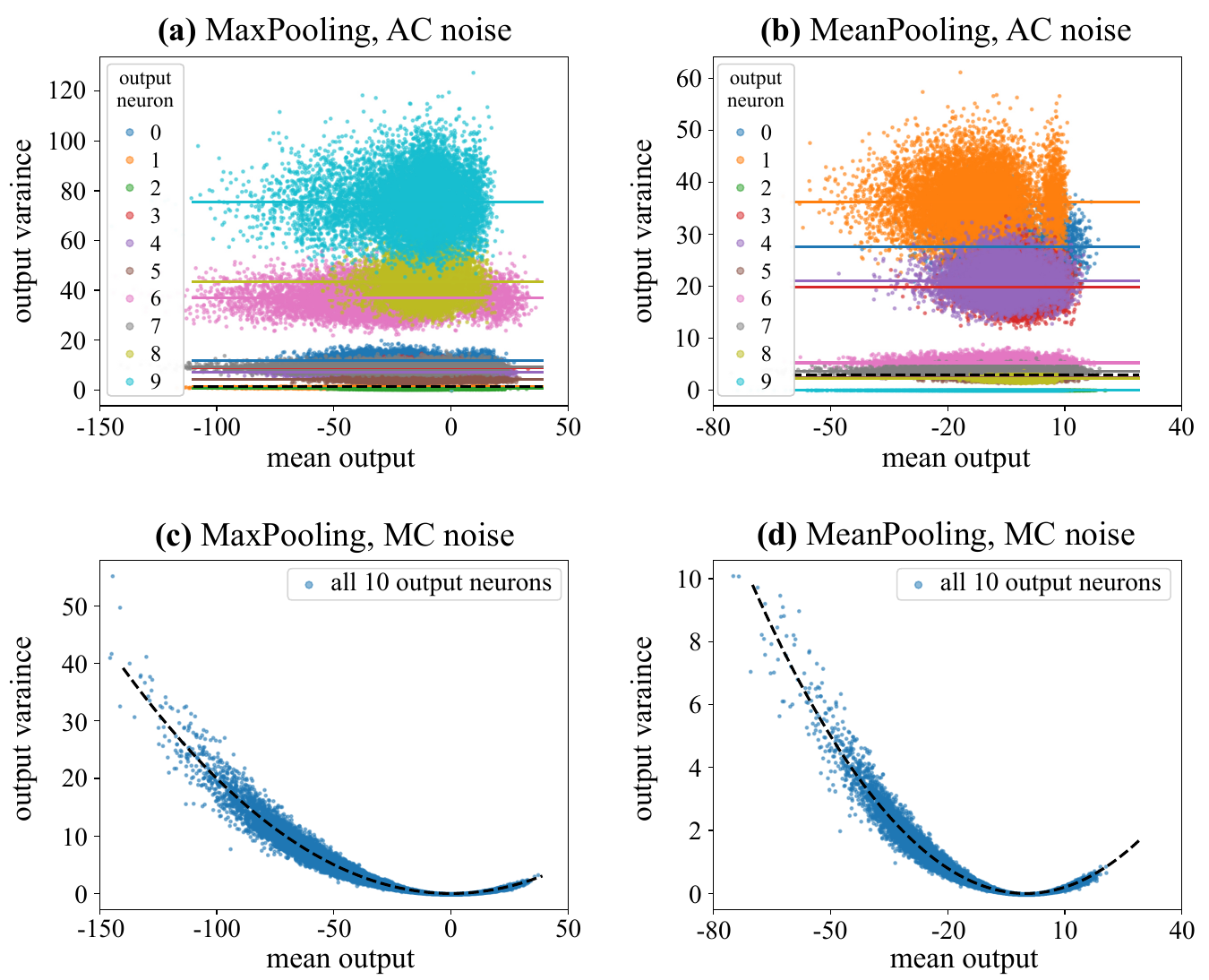}
\caption{\label{fig:pool_corr} The impact of correlated noise on the output of trained CNNs with MaxPooling (a,c) and MeanPooling (b,d). All panels contain the dependencies of variance (dispersion) on mean of the output signal from each of 10 output neurons. Considered noise types: (a,b) -- additive correlated noise with $D^C_A=10^{-3}$, (c,d) -- multiplicative correlated noise with $D^C_M=10^{-3}$. The results of numerical simulation are shown by points, while the lines of different colors correspond to analytical estimation of the noise level based on (\ref{eq:varXout_final2}) with averaging over one of 10 rows of $\mathbf{W}^\mathrm{out}$, black dashed lines were prepared using averaging over the entire matrix.}
\end{figure}

Figure \ref{fig:pool_corr} shows the impact of correlated noise in the same form as in Fig.~\ref{fig:pool_uncorr}. Our prediction based on (\ref{eq:varXout_final2}) works very well in this case. From this we can conclude that the use of pooling does not essentially affect the reduction of correlated noise. However, the use of pooling can indirectly lead to changes in the statistical characteristics of the connection matrices between neurons during the training process. For example, in Fig.~\ref{fig:pool_corr} the level of variance is much lower than in the similar Fig.~\ref{fig:conv_corr} without pooling.

In the case of additive correlated noise (Fig.~\ref{fig:pool_corr}(a,b)), it is clear that both methods of introducing a pooling layer can reduce the influence of noise. This result was obtained many times for different trained networks.

In the presence of multiplicative correlated noise in the convolutional layer (Fig.~\ref{fig:pool_corr}(c,d)), a significant improvement in network performance can only be achieved if MeanPooling is used. If the noise is multiplicative, then MaxPooling increases the output signal of CNN and therefore the final variance becomes larger than without MaxPooling.

\section{Conclusions}\label{sec:conclu}
In this paper, the impact of noise on a simplified trained convolutional network has been considered. Here we propose an analytical assessment of the noise level in CNN's output signal. It shows a good correlation with the results of numerical simulation. Four types of noise were considered: additive correlated noise, additive uncorrelated noise, multiplicative correlated noise, multiplicative uncorrelated noise. The following results were obtained for all types of noise:

1. The propagation of additive correlated and uncorrelated noise strongly depends on the statistical characteristics of the matrices. In the case of additive uncorrelated noise, the main important characteristic is the value of $N\eta(\mathbf{W}^\mathrm{out})$. Decreasing this value results in less additive uncorrelated noise. In the case of additive correlated noise, the same can be said about the value of $\Big(N\mu(\mathbf{W}^\mathrm{out})\Big)^2$. Sometimes it may turn out that the matrix $\mathbf{W}^\mathrm{out}$ is very non-uniform. This can lead to the fact that the noise enters the neurons of the next layer unevenly, resulting in the focusing of the output variance (dispersion) around several levels. If this is the case, analytical estimation of the noise level requires averaging of the above values not over the entire matrix, but over individual columns corresponding to the output neurons.

2. Using the obtained analytics, it was found that multiplicative correlated noise essentially does not depend on the statistical characteristics of the matrix. In the case of multiplicative correlated noise, the dispersion of the network output signal is directly proportional to the square of network's output signal and the noise intensity. However, it is important to maintain the output signal magnitude above the noise intensity to prevent the useful signal from being completely lost.

3. An analytical prediction of the noise level has also been proposed for multiplicative uncorrelated noise. However, it requires knowledge of the output signals of neurons from hidden layers, which is not always possible without numerical simulation.

The presence of a pooling layer after the convolutional layer greatly reduces the influence of noise. We have considered MaxPooling and MeanPooling. MeanPooling allows to greatly reduce the influence of uncorrelated noise. Correlated noise is not reduced by pooling, but its use leads to other connection matrices during training. The statistical properties of connection matrices in CNNs with and without pooling can be quite different, which indirectly leads to a change in the variance of the network output in the case of correlated noise. Using MaxPooling is not recommended in networks with any multiplicative noise. The variance of the output signal of a network with multiplicative noise is proportional to the square of the output signal, and its increase leads to a quadratic increase in variance.

%\section*{Supplementary Material}
%Our supplementary material contains a pdf-file with additional description of noise reduction methods and how they can be implemented using Python code. It starts with pooling technique and then we consider all three ghost neurons.

\begin{acknowledgments}
This work was supported by the Russian Science Foundation (project No. 23-72-01094). %\hyperref{https://rscf.ru/project/23-72-01094/}
\end{acknowledgments}

\section*{Data Availability Statement}
The data that support the findings of this study are available from the corresponding author upon reasonable request.

\section*{References}
%\bibliography{bibliography}% Produces the bibliography via BibTeX.

\begin{thebibliography}{25}%
\makeatletter
\providecommand \@ifxundefined [1]{%
 \@ifx{#1\undefined}
}%
\providecommand \@ifnum [1]{%
 \ifnum #1\expandafter \@firstoftwo
 \else \expandafter \@secondoftwo
 \fi
}%
\providecommand \@ifx [1]{%
 \ifx #1\expandafter \@firstoftwo
 \else \expandafter \@secondoftwo
 \fi
}%
\providecommand \natexlab [1]{#1}%
\providecommand \enquote  [1]{``#1''}%
\providecommand \bibnamefont  [1]{#1}%
\providecommand \bibfnamefont [1]{#1}%
\providecommand \citenamefont [1]{#1}%
\providecommand \href@noop [0]{\@secondoftwo}%
\providecommand \href [0]{\begingroup \@sanitize@url \@href}%
\providecommand \@href[1]{\@@startlink{#1}\@@href}%
\providecommand \@@href[1]{\endgroup#1\@@endlink}%
\providecommand \@sanitize@url [0]{\catcode `\\12\catcode `\$12\catcode
  `\&12\catcode `\#12\catcode `\^12\catcode `\_12\catcode `\%12\relax}%
\providecommand \@@startlink[1]{}%
\providecommand \@@endlink[0]{}%
\providecommand \url  [0]{\begingroup\@sanitize@url \@url }%
\providecommand \@url [1]{\endgroup\@href {#1}{\urlprefix }}%
\providecommand \urlprefix  [0]{URL }%
\providecommand \Eprint [0]{\href }%
\providecommand \doibase [0]{http://dx.doi.org/}%
\providecommand \selectlanguage [0]{\@gobble}%
\providecommand \bibinfo  [0]{\@secondoftwo}%
\providecommand \bibfield  [0]{\@secondoftwo}%
\providecommand \translation [1]{[#1]}%
\providecommand \BibitemOpen [0]{}%
\providecommand \bibitemStop [0]{}%
\providecommand \bibitemNoStop [0]{.\EOS\space}%
\providecommand \EOS [0]{\spacefactor3000\relax}%
\providecommand \BibitemShut  [1]{\csname bibitem#1\endcsname}%
\let\auto@bib@innerbib\@empty
%</preamble>
\bibitem [{\citenamefont {Karniadakis}\ \emph {et~al.}(2021)\citenamefont
  {Karniadakis}, \citenamefont {Kevrekidis}, \citenamefont {Lu}, \citenamefont
  {Perdikaris}, \citenamefont {Wang},\ and\ \citenamefont
  {Yang}}]{Karniadakis2021}%
  \BibitemOpen
  \bibfield  {author} {\bibinfo {author} {\bibfnamefont {G.~E.}\ \bibnamefont
  {Karniadakis}}, \bibinfo {author} {\bibfnamefont {I.~G.}\ \bibnamefont
  {Kevrekidis}}, \bibinfo {author} {\bibfnamefont {L.}~\bibnamefont {Lu}},
  \bibinfo {author} {\bibfnamefont {P.}~\bibnamefont {Perdikaris}}, \bibinfo
  {author} {\bibfnamefont {S.}~\bibnamefont {Wang}}, \ and\ \bibinfo {author}
  {\bibfnamefont {L.}~\bibnamefont {Yang}},\ }\bibfield  {title} {\enquote
  {\bibinfo {title} {Physics-informed machine learning},}\ }\href {\doibase
  10.1038/s42254-021-00314-5} {\bibfield  {journal} {\bibinfo  {journal}
  {Nature Reviews Physics}\ }\textbf {\bibinfo {volume} {3}},\ \bibinfo {pages}
  {422--440} (\bibinfo {year} {2021})}\BibitemShut {NoStop}%
\bibitem [{\citenamefont {LeCun}, \citenamefont {Bengio},\ and\ \citenamefont
  {Hinton}(2015)}]{Lecun2015}%
  \BibitemOpen
  \bibfield  {author} {\bibinfo {author} {\bibfnamefont {Y.}~\bibnamefont
  {LeCun}}, \bibinfo {author} {\bibfnamefont {Y.}~\bibnamefont {Bengio}}, \
  and\ \bibinfo {author} {\bibfnamefont {G.}~\bibnamefont {Hinton}},\
  }\bibfield  {title} {\enquote {\bibinfo {title} {Deep learning},}\ }\href
  {\doibase 10.1038/nature14539} {\bibfield  {journal} {\bibinfo  {journal}
  {Nature}\ }\textbf {\bibinfo {volume} {521}},\ \bibinfo {pages} {436--444}
  (\bibinfo {year} {2015})}\BibitemShut {NoStop}%
\bibitem [{\citenamefont {Krizhevsky}, \citenamefont {Sutskever},\ and\
  \citenamefont {Hinton}(2017)}]{Krizhevsky2017}%
  \BibitemOpen
  \bibfield  {author} {\bibinfo {author} {\bibfnamefont {A.}~\bibnamefont
  {Krizhevsky}}, \bibinfo {author} {\bibfnamefont {I.}~\bibnamefont
  {Sutskever}}, \ and\ \bibinfo {author} {\bibfnamefont {G.~E.}\ \bibnamefont
  {Hinton}},\ }\bibfield  {title} {\enquote {\bibinfo {title} {Imagenet
  classification with deep convolutional neural networks},}\ }\href {\doibase
  10.1145/3065386} {\bibfield  {journal} {\bibinfo  {journal} {Commun. ACM}\
  }\textbf {\bibinfo {volume} {60}},\ \bibinfo {pages} {84--90} (\bibinfo
  {year} {2017})}\BibitemShut {NoStop}%
\bibitem [{\citenamefont {Maturana}\ and\ \citenamefont
  {Scherer}(2015)}]{Maturana2015}%
  \BibitemOpen
  \bibfield  {author} {\bibinfo {author} {\bibfnamefont {D.}~\bibnamefont
  {Maturana}}\ and\ \bibinfo {author} {\bibfnamefont {S.}~\bibnamefont
  {Scherer}},\ }\bibfield  {title} {\enquote {\bibinfo {title} {Voxnet: A 3d
  convolutional neural network for real-time object recognition},}\ }in\ \href
  {\doibase 10.1109/IROS.2015.7353481} {\emph {\bibinfo {booktitle} {2015
  IEEE/RSJ International Conference on Intelligent Robots and Systems
  (IROS)}}}\ (\bibinfo {year} {2015})\ pp.\ \bibinfo {pages}
  {922--928}\BibitemShut {NoStop}%
\bibitem [{\citenamefont {Graves}, \citenamefont {Mohamed},\ and\ \citenamefont
  {Hinton}(2013)}]{Graves2013}%
  \BibitemOpen
  \bibfield  {author} {\bibinfo {author} {\bibfnamefont {A.}~\bibnamefont
  {Graves}}, \bibinfo {author} {\bibfnamefont {A.-r.}\ \bibnamefont {Mohamed}},
  \ and\ \bibinfo {author} {\bibfnamefont {G.}~\bibnamefont {Hinton}},\
  }\bibfield  {title} {\enquote {\bibinfo {title} {Speech recognition with deep
  recurrent neural networks},}\ }in\ \href {\doibase
  10.1109/ICASSP.2013.6638947} {\emph {\bibinfo {booktitle} {2013 IEEE
  International Conference on Acoustics, Speech and Signal Processing}}}\
  (\bibinfo {year} {2013})\ pp.\ \bibinfo {pages} {6645--6649}\BibitemShut
  {NoStop}%
\bibitem [{\citenamefont {Kar}\ and\ \citenamefont {Moura}(2009)}]{Kar2009}%
  \BibitemOpen
  \bibfield  {author} {\bibinfo {author} {\bibfnamefont {S.}~\bibnamefont
  {Kar}}\ and\ \bibinfo {author} {\bibfnamefont {J.~M.~F.}\ \bibnamefont
  {Moura}},\ }\bibfield  {title} {\enquote {\bibinfo {title} {Distributed
  consensus algorithms in sensor networks with imperfect communication: Link
  failures and channel noise},}\ }\href {\doibase 10.1109/TSP.2008.2007111}
  {\bibfield  {journal} {\bibinfo  {journal} {IEEE Transactions on Signal
  Processing}\ }\textbf {\bibinfo {volume} {57}},\ \bibinfo {pages} {355--369}
  (\bibinfo {year} {2009})}\BibitemShut {NoStop}%
\bibitem [{\citenamefont {Aguirre}\ \emph {et~al.}(2024)\citenamefont
  {Aguirre}, \citenamefont {Sebastian}, \citenamefont {Le~Gallo}, \citenamefont
  {Song}, \citenamefont {Wang}, \citenamefont {Yang}, \citenamefont {Lu},
  \citenamefont {Chang}, \citenamefont {Ielmini}, \citenamefont {Yang},
  \citenamefont {Mehonic}, \citenamefont {Kenyon}, \citenamefont {Villena},
  \citenamefont {Rold{\'a}n}, \citenamefont {Wu}, \citenamefont {Hsu},
  \citenamefont {Raghavan}, \citenamefont {Su{\~n}{\'e}}, \citenamefont
  {Miranda}, \citenamefont {Eltawil}, \citenamefont {Setti}, \citenamefont
  {Smagulova}, \citenamefont {Salama}, \citenamefont {Krestinskaya},
  \citenamefont {Yan}, \citenamefont {Ang}, \citenamefont {Jain}, \citenamefont
  {Li}, \citenamefont {Alharbi}, \citenamefont {Pazos},\ and\ \citenamefont
  {Lanza}}]{Aguirre2024}%
  \BibitemOpen
  \bibfield  {author} {\bibinfo {author} {\bibfnamefont {F.}~\bibnamefont
  {Aguirre}}, \bibinfo {author} {\bibfnamefont {A.}~\bibnamefont {Sebastian}},
  \bibinfo {author} {\bibfnamefont {M.}~\bibnamefont {Le~Gallo}}, \bibinfo
  {author} {\bibfnamefont {W.}~\bibnamefont {Song}}, \bibinfo {author}
  {\bibfnamefont {T.}~\bibnamefont {Wang}}, \bibinfo {author} {\bibfnamefont
  {J.~J.}\ \bibnamefont {Yang}}, \bibinfo {author} {\bibfnamefont
  {W.}~\bibnamefont {Lu}}, \bibinfo {author} {\bibfnamefont {M.-F.}\
  \bibnamefont {Chang}}, \bibinfo {author} {\bibfnamefont {D.}~\bibnamefont
  {Ielmini}}, \bibinfo {author} {\bibfnamefont {Y.}~\bibnamefont {Yang}},
  \bibinfo {author} {\bibfnamefont {A.}~\bibnamefont {Mehonic}}, \bibinfo
  {author} {\bibfnamefont {A.}~\bibnamefont {Kenyon}}, \bibinfo {author}
  {\bibfnamefont {M.~A.}\ \bibnamefont {Villena}}, \bibinfo {author}
  {\bibfnamefont {J.~B.}\ \bibnamefont {Rold{\'a}n}}, \bibinfo {author}
  {\bibfnamefont {Y.}~\bibnamefont {Wu}}, \bibinfo {author} {\bibfnamefont
  {H.-H.}\ \bibnamefont {Hsu}}, \bibinfo {author} {\bibfnamefont
  {N.}~\bibnamefont {Raghavan}}, \bibinfo {author} {\bibfnamefont
  {J.}~\bibnamefont {Su{\~n}{\'e}}}, \bibinfo {author} {\bibfnamefont
  {E.}~\bibnamefont {Miranda}}, \bibinfo {author} {\bibfnamefont
  {A.}~\bibnamefont {Eltawil}}, \bibinfo {author} {\bibfnamefont
  {G.}~\bibnamefont {Setti}}, \bibinfo {author} {\bibfnamefont
  {K.}~\bibnamefont {Smagulova}}, \bibinfo {author} {\bibfnamefont {K.~N.}\
  \bibnamefont {Salama}}, \bibinfo {author} {\bibfnamefont {O.}~\bibnamefont
  {Krestinskaya}}, \bibinfo {author} {\bibfnamefont {X.}~\bibnamefont {Yan}},
  \bibinfo {author} {\bibfnamefont {K.-W.}\ \bibnamefont {Ang}}, \bibinfo
  {author} {\bibfnamefont {S.}~\bibnamefont {Jain}}, \bibinfo {author}
  {\bibfnamefont {S.}~\bibnamefont {Li}}, \bibinfo {author} {\bibfnamefont
  {O.}~\bibnamefont {Alharbi}}, \bibinfo {author} {\bibfnamefont
  {S.}~\bibnamefont {Pazos}}, \ and\ \bibinfo {author} {\bibfnamefont
  {M.}~\bibnamefont {Lanza}},\ }\bibfield  {title} {\enquote {\bibinfo {title}
  {Hardware implementation of memristor-based artificial neural networks},}\
  }\href {\doibase 10.1038/s41467-024-45670-9} {\bibfield  {journal} {\bibinfo
  {journal} {Nature Communications}\ }\textbf {\bibinfo {volume} {15}},\
  \bibinfo {pages} {1974} (\bibinfo {year} {2024})}\BibitemShut {NoStop}%
\bibitem [{\citenamefont {Chen}\ \emph {et~al.}(2023)\citenamefont {Chen},
  \citenamefont {Nazhamaiti}, \citenamefont {Xu}, \citenamefont {Meng},
  \citenamefont {Zhou}, \citenamefont {Li}, \citenamefont {Fan}, \citenamefont
  {Wei}, \citenamefont {Wu}, \citenamefont {Qiao}, \citenamefont {Fang},\ and\
  \citenamefont {Dai}}]{Chen2023}%
  \BibitemOpen
  \bibfield  {author} {\bibinfo {author} {\bibfnamefont {Y.}~\bibnamefont
  {Chen}}, \bibinfo {author} {\bibfnamefont {M.}~\bibnamefont {Nazhamaiti}},
  \bibinfo {author} {\bibfnamefont {H.}~\bibnamefont {Xu}}, \bibinfo {author}
  {\bibfnamefont {Y.}~\bibnamefont {Meng}}, \bibinfo {author} {\bibfnamefont
  {T.}~\bibnamefont {Zhou}}, \bibinfo {author} {\bibfnamefont {G.}~\bibnamefont
  {Li}}, \bibinfo {author} {\bibfnamefont {J.}~\bibnamefont {Fan}}, \bibinfo
  {author} {\bibfnamefont {Q.}~\bibnamefont {Wei}}, \bibinfo {author}
  {\bibfnamefont {J.}~\bibnamefont {Wu}}, \bibinfo {author} {\bibfnamefont
  {F.}~\bibnamefont {Qiao}}, \bibinfo {author} {\bibfnamefont {L.}~\bibnamefont
  {Fang}}, \ and\ \bibinfo {author} {\bibfnamefont {Q.}~\bibnamefont {Dai}},\
  }\bibfield  {title} {\enquote {\bibinfo {title} {All-analog photoelectronic
  chip for high-speed vision tasks},}\ }\href {\doibase
  10.1038/s41586-023-06558-8} {\bibfield  {journal} {\bibinfo  {journal}
  {Nature}\ }\textbf {\bibinfo {volume} {623}},\ \bibinfo {pages} {48--57}
  (\bibinfo {year} {2023})}\BibitemShut {NoStop}%
\bibitem [{\citenamefont {Brunner}\ \emph {et~al.}(2013)\citenamefont
  {Brunner}, \citenamefont {Soriano}, \citenamefont {Mirasso},\ and\
  \citenamefont {Fischer}}]{Brunner2013a}%
  \BibitemOpen
  \bibfield  {author} {\bibinfo {author} {\bibfnamefont {D.}~\bibnamefont
  {Brunner}}, \bibinfo {author} {\bibfnamefont {M.~C.}\ \bibnamefont
  {Soriano}}, \bibinfo {author} {\bibfnamefont {C.~R.}\ \bibnamefont
  {Mirasso}}, \ and\ \bibinfo {author} {\bibfnamefont {I.}~\bibnamefont
  {Fischer}},\ }\bibfield  {title} {\enquote {\bibinfo {title} {{Parallel
  photonic information processing at gigabyte per second data rates using
  transient states}},}\ }\href {\doibase 10.1038/ncomms2368} {\bibfield
  {journal} {\bibinfo  {journal} {Nature communications}\ }\textbf {\bibinfo
  {volume} {4}},\ \bibinfo {pages} {1364} (\bibinfo {year} {2013})}\BibitemShut
  {NoStop}%
\bibitem [{\citenamefont {Tuma}\ \emph {et~al.}(2016)\citenamefont {Tuma},
  \citenamefont {Pantazi}, \citenamefont {{Le Gallo}}, \citenamefont
  {Sebastian},\ and\ \citenamefont {Eleftheriou}}]{Tuma2016}%
  \BibitemOpen
  \bibfield  {author} {\bibinfo {author} {\bibfnamefont {T.}~\bibnamefont
  {Tuma}}, \bibinfo {author} {\bibfnamefont {A.}~\bibnamefont {Pantazi}},
  \bibinfo {author} {\bibfnamefont {M.}~\bibnamefont {{Le Gallo}}}, \bibinfo
  {author} {\bibfnamefont {A.}~\bibnamefont {Sebastian}}, \ and\ \bibinfo
  {author} {\bibfnamefont {E.}~\bibnamefont {Eleftheriou}},\ }\bibfield
  {title} {\enquote {\bibinfo {title} {{Stochastic phase-change neurons}},}\
  }\href {\doibase 10.1038/nnano.2016.70} {\bibfield  {journal} {\bibinfo
  {journal} {Nature Nanotechnology}\ }\textbf {\bibinfo {volume} {11}},\
  \bibinfo {pages} {693--699} (\bibinfo {year} {2016})}\BibitemShut {NoStop}%
\bibitem [{\citenamefont {Torrejon}\ \emph {et~al.}(2017)\citenamefont
  {Torrejon}, \citenamefont {Riou}, \citenamefont {Araujo}, \citenamefont
  {Tsunegi}, \citenamefont {Khalsa}, \citenamefont {Querlioz}, \citenamefont
  {Bortolotti}, \citenamefont {Cros}, \citenamefont {Yakushiji}, \citenamefont
  {Fukushima}, \citenamefont {Kubota}, \citenamefont {Yuasa}, \citenamefont
  {Stiles},\ and\ \citenamefont {Grollier}}]{Torrejon2017}%
  \BibitemOpen
  \bibfield  {author} {\bibinfo {author} {\bibfnamefont {J.}~\bibnamefont
  {Torrejon}}, \bibinfo {author} {\bibfnamefont {M.}~\bibnamefont {Riou}},
  \bibinfo {author} {\bibfnamefont {F.~A.}\ \bibnamefont {Araujo}}, \bibinfo
  {author} {\bibfnamefont {S.}~\bibnamefont {Tsunegi}}, \bibinfo {author}
  {\bibfnamefont {G.}~\bibnamefont {Khalsa}}, \bibinfo {author} {\bibfnamefont
  {D.}~\bibnamefont {Querlioz}}, \bibinfo {author} {\bibfnamefont
  {P.}~\bibnamefont {Bortolotti}}, \bibinfo {author} {\bibfnamefont
  {V.}~\bibnamefont {Cros}}, \bibinfo {author} {\bibfnamefont {K.}~\bibnamefont
  {Yakushiji}}, \bibinfo {author} {\bibfnamefont {A.}~\bibnamefont
  {Fukushima}}, \bibinfo {author} {\bibfnamefont {H.}~\bibnamefont {Kubota}},
  \bibinfo {author} {\bibfnamefont {S.}~\bibnamefont {Yuasa}}, \bibinfo
  {author} {\bibfnamefont {M.~D.}\ \bibnamefont {Stiles}}, \ and\ \bibinfo
  {author} {\bibfnamefont {J.}~\bibnamefont {Grollier}},\ }\bibfield  {title}
  {\enquote {\bibinfo {title} {Neuromorphic computing with nanoscale spintronic
  oscillators},}\ }\href {\doibase 10.1038/nature23011} {\bibfield  {journal}
  {\bibinfo  {journal} {Nature}\ }\textbf {\bibinfo {volume} {547}},\ \bibinfo
  {pages} {428--431} (\bibinfo {year} {2017})}\BibitemShut {NoStop}%
\bibitem [{\citenamefont {Psaltis}\ \emph {et~al.}(1990)\citenamefont
  {Psaltis}, \citenamefont {Brady}, \citenamefont {Gu},\ and\ \citenamefont
  {Lin}}]{Psaltis1990}%
  \BibitemOpen
  \bibfield  {author} {\bibinfo {author} {\bibfnamefont {D.}~\bibnamefont
  {Psaltis}}, \bibinfo {author} {\bibfnamefont {D.}~\bibnamefont {Brady}},
  \bibinfo {author} {\bibfnamefont {X.-G.}\ \bibnamefont {Gu}}, \ and\ \bibinfo
  {author} {\bibfnamefont {S.}~\bibnamefont {Lin}},\ }\bibfield  {title}
  {\enquote {\bibinfo {title} {{Holography in artificial neural networks}},}\
  }\href {\doibase 10.1038/343325a0} {\bibfield  {journal} {\bibinfo  {journal}
  {Nature}\ }\textbf {\bibinfo {volume} {343}},\ \bibinfo {pages} {325--330}
  (\bibinfo {year} {1990})}\BibitemShut {NoStop}%
\bibitem [{\citenamefont {Bueno}\ \emph {et~al.}(2018)\citenamefont {Bueno},
  \citenamefont {Maktoobi}, \citenamefont {Froehly}, \citenamefont {Fischer},
  \citenamefont {Jacquot}, \citenamefont {Larger},\ and\ \citenamefont
  {Brunner}}]{Bueno2018}%
  \BibitemOpen
  \bibfield  {author} {\bibinfo {author} {\bibfnamefont {J.}~\bibnamefont
  {Bueno}}, \bibinfo {author} {\bibfnamefont {S.}~\bibnamefont {Maktoobi}},
  \bibinfo {author} {\bibfnamefont {L.}~\bibnamefont {Froehly}}, \bibinfo
  {author} {\bibfnamefont {I.}~\bibnamefont {Fischer}}, \bibinfo {author}
  {\bibfnamefont {M.}~\bibnamefont {Jacquot}}, \bibinfo {author} {\bibfnamefont
  {L.}~\bibnamefont {Larger}}, \ and\ \bibinfo {author} {\bibfnamefont
  {D.}~\bibnamefont {Brunner}},\ }\bibfield  {title} {\enquote {\bibinfo
  {title} {{Reinforcement Learning in a large scale photonic Recurrent Neural
  Network}},}\ }\href {\doibase 10.1364/OPTICA.5.000756} {\bibfield  {journal}
  {\bibinfo  {journal} {Optica}\ }\textbf {\bibinfo {volume} {5}},\ \bibinfo
  {pages} {756 -- 760} (\bibinfo {year} {2018})}\BibitemShut {NoStop}%
\bibitem [{\citenamefont {Lin}\ \emph {et~al.}(2018)\citenamefont {Lin},
  \citenamefont {Rivenson}, \citenamefont {Yardimci}, \citenamefont {Veli},
  \citenamefont {Jarrahi},\ and\ \citenamefont {Ozcan}}]{Lin2018}%
  \BibitemOpen
  \bibfield  {author} {\bibinfo {author} {\bibfnamefont {X.}~\bibnamefont
  {Lin}}, \bibinfo {author} {\bibfnamefont {Y.}~\bibnamefont {Rivenson}},
  \bibinfo {author} {\bibfnamefont {N.~T.}\ \bibnamefont {Yardimci}}, \bibinfo
  {author} {\bibfnamefont {M.}~\bibnamefont {Veli}}, \bibinfo {author}
  {\bibfnamefont {M.}~\bibnamefont {Jarrahi}}, \ and\ \bibinfo {author}
  {\bibfnamefont {A.}~\bibnamefont {Ozcan}},\ }\bibfield  {title} {\enquote
  {\bibinfo {title} {{All-Optical Machine Learning Using Diffractive Deep
  Neural Networks}},}\ }\href {\doibase 10.1126/science.aat8084} {\bibfield
  {journal} {\bibinfo  {journal} {Science}\ }\textbf {\bibinfo {volume}
  {361}},\ \bibinfo {pages} {1004--1008} (\bibinfo {year} {2018})}\BibitemShut
  {NoStop}%
\bibitem [{\citenamefont {Shen}\ \emph {et~al.}(2017)\citenamefont {Shen},
  \citenamefont {Harris}, \citenamefont {Skirlo}, \citenamefont {Prabhu},
  \citenamefont {Baehr-Jones}, \citenamefont {Hochberg}, \citenamefont {Sun},
  \citenamefont {Zhao}, \citenamefont {Larochelle}, \citenamefont {Englund},\
  and\ \citenamefont {Soljacic}}]{Shen2017}%
  \BibitemOpen
  \bibfield  {author} {\bibinfo {author} {\bibfnamefont {Y.}~\bibnamefont
  {Shen}}, \bibinfo {author} {\bibfnamefont {N.~C.}\ \bibnamefont {Harris}},
  \bibinfo {author} {\bibfnamefont {S.}~\bibnamefont {Skirlo}}, \bibinfo
  {author} {\bibfnamefont {M.}~\bibnamefont {Prabhu}}, \bibinfo {author}
  {\bibfnamefont {T.}~\bibnamefont {Baehr-Jones}}, \bibinfo {author}
  {\bibfnamefont {M.}~\bibnamefont {Hochberg}}, \bibinfo {author}
  {\bibfnamefont {X.}~\bibnamefont {Sun}}, \bibinfo {author} {\bibfnamefont
  {S.}~\bibnamefont {Zhao}}, \bibinfo {author} {\bibfnamefont {H.}~\bibnamefont
  {Larochelle}}, \bibinfo {author} {\bibfnamefont {D.}~\bibnamefont {Englund}},
  \ and\ \bibinfo {author} {\bibfnamefont {M.}~\bibnamefont {Soljacic}},\
  }\bibfield  {title} {\enquote {\bibinfo {title} {{Deep Learning with Coherent
  Nanophotonic Circuits}},}\ }\href {\doibase 10.1038/nphoton.2017.93}
  {\bibfield  {journal} {\bibinfo  {journal} {Nature Photonics}\ }\textbf
  {\bibinfo {volume} {11}},\ \bibinfo {pages} {441--446} (\bibinfo {year}
  {2017})}\BibitemShut {NoStop}%
\bibitem [{\citenamefont {Tait}\ \emph {et~al.}(2017)\citenamefont {Tait},
  \citenamefont {{De Lima}}, \citenamefont {Zhou}, \citenamefont {Wu},
  \citenamefont {Nahmias}, \citenamefont {Shastri},\ and\ \citenamefont
  {Prucnal}}]{Tait2017}%
  \BibitemOpen
  \bibfield  {author} {\bibinfo {author} {\bibfnamefont {A.~N.}\ \bibnamefont
  {Tait}}, \bibinfo {author} {\bibfnamefont {T.~F.}\ \bibnamefont {{De Lima}}},
  \bibinfo {author} {\bibfnamefont {E.}~\bibnamefont {Zhou}}, \bibinfo {author}
  {\bibfnamefont {A.~X.}\ \bibnamefont {Wu}}, \bibinfo {author} {\bibfnamefont
  {M.~A.}\ \bibnamefont {Nahmias}}, \bibinfo {author} {\bibfnamefont {B.~J.}\
  \bibnamefont {Shastri}}, \ and\ \bibinfo {author} {\bibfnamefont {P.~R.}\
  \bibnamefont {Prucnal}},\ }\bibfield  {title} {\enquote {\bibinfo {title}
  {{Neuromorphic photonic networks using silicon photonic weight banks}},}\
  }\href {\doibase 10.1038/s41598-017-07754-z} {\bibfield  {journal} {\bibinfo
  {journal} {Scientific Reports}\ }\textbf {\bibinfo {volume} {7}},\ \bibinfo
  {pages} {7430} (\bibinfo {year} {2017})}\BibitemShut {NoStop}%
\bibitem [{\citenamefont {Moughames}\ \emph
  {et~al.}(2020{\natexlab{a}})\citenamefont {Moughames}, \citenamefont {Porte},
  \citenamefont {Thiel}, \citenamefont {Ulliac}, \citenamefont {Larger},
  \citenamefont {Jacquot}, \citenamefont {Kadic},\ and\ \citenamefont
  {Brunner}}]{Moughames2020}%
  \BibitemOpen
  \bibfield  {author} {\bibinfo {author} {\bibfnamefont {J.}~\bibnamefont
  {Moughames}}, \bibinfo {author} {\bibfnamefont {X.}~\bibnamefont {Porte}},
  \bibinfo {author} {\bibfnamefont {M.}~\bibnamefont {Thiel}}, \bibinfo
  {author} {\bibfnamefont {G.}~\bibnamefont {Ulliac}}, \bibinfo {author}
  {\bibfnamefont {L.}~\bibnamefont {Larger}}, \bibinfo {author} {\bibfnamefont
  {M.}~\bibnamefont {Jacquot}}, \bibinfo {author} {\bibfnamefont
  {M.}~\bibnamefont {Kadic}}, \ and\ \bibinfo {author} {\bibfnamefont
  {D.}~\bibnamefont {Brunner}},\ }\bibfield  {title} {\enquote {\bibinfo
  {title} {Three-dimensional waveguide interconnects for scalable integration
  of photonic neural networks},}\ }\href {\doibase 10.1364/OPTICA.388205}
  {\bibfield  {journal} {\bibinfo  {journal} {Optica}\ }\textbf {\bibinfo
  {volume} {7}},\ \bibinfo {pages} {640--646} (\bibinfo {year}
  {2020}{\natexlab{a}})}\BibitemShut {NoStop}%
\bibitem [{\citenamefont {{Dinc, Niyazi Ulas}}, \citenamefont {{Psaltis,
  Demetri}},\ and\ \citenamefont {{Brunner, Daniel}}(2020)}]{Dinc2020}%
  \BibitemOpen
  \bibfield  {author} {\bibinfo {author} {\bibnamefont {{Dinc, Niyazi Ulas}}},
  \bibinfo {author} {\bibnamefont {{Psaltis, Demetri}}}, \ and\ \bibinfo
  {author} {\bibnamefont {{Brunner, Daniel}}},\ }\bibfield  {title} {\enquote
  {\bibinfo {title} {Optical neural networks: The 3d connection},}\ }\href
  {\doibase 10.1051/photon/202010434} {\bibfield  {journal} {\bibinfo
  {journal} {Photoniques}\ ,\ \bibinfo {pages} {34--38}} (\bibinfo {year}
  {2020})}\BibitemShut {NoStop}%
\bibitem [{\citenamefont {Moughames}\ \emph
  {et~al.}(2020{\natexlab{b}})\citenamefont {Moughames}, \citenamefont {Porte},
  \citenamefont {Larger}, \citenamefont {Jacquot}, \citenamefont {Kadic},\ and\
  \citenamefont {Brunner}}]{Moughames2020a}%
  \BibitemOpen
  \bibfield  {author} {\bibinfo {author} {\bibfnamefont {J.}~\bibnamefont
  {Moughames}}, \bibinfo {author} {\bibfnamefont {X.}~\bibnamefont {Porte}},
  \bibinfo {author} {\bibfnamefont {L.}~\bibnamefont {Larger}}, \bibinfo
  {author} {\bibfnamefont {M.}~\bibnamefont {Jacquot}}, \bibinfo {author}
  {\bibfnamefont {M.}~\bibnamefont {Kadic}}, \ and\ \bibinfo {author}
  {\bibfnamefont {D.}~\bibnamefont {Brunner}},\ }\bibfield  {title} {\enquote
  {\bibinfo {title} {3d printed multimode-splitters for photonic
  interconnects},}\ }\href {\doibase 10.1364/OME.402974} {\bibfield  {journal}
  {\bibinfo  {journal} {Opt. Mater. Express}\ }\textbf {\bibinfo {volume}
  {10}},\ \bibinfo {pages} {2952--2961} (\bibinfo {year}
  {2020}{\natexlab{b}})}\BibitemShut {NoStop}%
\bibitem [{\citenamefont {Semenova}, \citenamefont {Larger},\ and\
  \citenamefont {Brunner}(2022)}]{Semenova2022NN}%
  \BibitemOpen
  \bibfield  {author} {\bibinfo {author} {\bibfnamefont {N.}~\bibnamefont
  {Semenova}}, \bibinfo {author} {\bibfnamefont {L.}~\bibnamefont {Larger}}, \
  and\ \bibinfo {author} {\bibfnamefont {D.}~\bibnamefont {Brunner}},\
  }\bibfield  {title} {\enquote {\bibinfo {title} {Understanding and mitigating
  noise in trained deep neural networks},}\ }\href {\doibase
  10.1016/j.neunet.2021.11.008} {\bibfield  {journal} {\bibinfo  {journal}
  {Neural Networks}\ }\textbf {\bibinfo {volume} {146}},\ \bibinfo {pages}
  {151--160} (\bibinfo {year} {2022})}\BibitemShut {NoStop}%
\bibitem [{\citenamefont {Semenova}(2024)}]{Semenova2024echo}%
  \BibitemOpen
  \bibfield  {author} {\bibinfo {author} {\bibfnamefont {N.}~\bibnamefont
  {Semenova}},\ }\bibfield  {title} {\enquote {\bibinfo {title} {Impact of
  white gaussian internal noise on analog echo-state neural networks},}\
  }\href@noop {} {\bibfield  {journal} {\bibinfo  {journal} {arXiv preprint
  arXiv:2405.07670}\ } (\bibinfo {year} {2024})}\BibitemShut {NoStop}%
\bibitem [{\citenamefont {Semenova}\ and\ \citenamefont
  {Brunner}(2022)}]{Semenova2022Chaos}%
  \BibitemOpen
  \bibfield  {author} {\bibinfo {author} {\bibfnamefont {N.}~\bibnamefont
  {Semenova}}\ and\ \bibinfo {author} {\bibfnamefont {D.}~\bibnamefont
  {Brunner}},\ }\bibfield  {title} {\enquote {\bibinfo {title}
  {{Noise-mitigation strategies in physical feedforward neural networks}},}\
  }\href {\doibase 10.1063/5.0096637} {\bibfield  {journal} {\bibinfo
  {journal} {Chaos: An Interdisciplinary Journal of Nonlinear Science}\
  }\textbf {\bibinfo {volume} {32}},\ \bibinfo {pages} {061106} (\bibinfo
  {year} {2022})}\BibitemShut {NoStop}%
\bibitem [{\citenamefont {Semenova}\ and\ \citenamefont
  {Brunner}(2024)}]{Semenova2024Chaos}%
  \BibitemOpen
  \bibfield  {author} {\bibinfo {author} {\bibfnamefont {N.}~\bibnamefont
  {Semenova}}\ and\ \bibinfo {author} {\bibfnamefont {D.}~\bibnamefont
  {Brunner}},\ }\bibfield  {title} {\enquote {\bibinfo {title} {Impact of white
  noise in artificial neural networks trained for classification: Performance
  and noise mitigation strategies},}\ }\href {\doibase 10.1063/5.0206807}
  {\bibfield  {journal} {\bibinfo  {journal} {Chaos: An Interdisciplinary
  Journal of Nonlinear Science}\ }\textbf {\bibinfo {volume} {34}},\ \bibinfo
  {pages} {051101} (\bibinfo {year} {2024})}\BibitemShut {NoStop}%
\bibitem [{\citenamefont {Li}\ \emph {et~al.}(2022)\citenamefont {Li},
  \citenamefont {Liu}, \citenamefont {Yang}, \citenamefont {Peng},\ and\
  \citenamefont {Zhou}}]{Li2022}%
  \BibitemOpen
  \bibfield  {author} {\bibinfo {author} {\bibfnamefont {Z.}~\bibnamefont
  {Li}}, \bibinfo {author} {\bibfnamefont {F.}~\bibnamefont {Liu}}, \bibinfo
  {author} {\bibfnamefont {W.}~\bibnamefont {Yang}}, \bibinfo {author}
  {\bibfnamefont {S.}~\bibnamefont {Peng}}, \ and\ \bibinfo {author}
  {\bibfnamefont {J.}~\bibnamefont {Zhou}},\ }\bibfield  {title} {\enquote
  {\bibinfo {title} {A survey of convolutional neural networks: Analysis,
  applications, and prospects},}\ }\href {\doibase 10.1109/TNNLS.2021.3084827}
  {\bibfield  {journal} {\bibinfo  {journal} {IEEE Transactions on Neural
  Networks and Learning Systems}\ }\textbf {\bibinfo {volume} {33}},\ \bibinfo
  {pages} {6999--7019} (\bibinfo {year} {2022})}\BibitemShut {NoStop}%
\bibitem [{\citenamefont {LeCun}(1998)}]{LeCun1998}%
  \BibitemOpen
  \bibfield  {author} {\bibinfo {author} {\bibfnamefont {Y.}~\bibnamefont
  {LeCun}},\ }\bibfield  {title} {\enquote {\bibinfo {title} {The mnist
  database of handwritten digits},}\ }\href {http://yann.lecun.com/exdb/mnist/}
  {\  (\bibinfo {year} {1998})}\BibitemShut {NoStop}%
\end{thebibliography}

\end{document}